\documentclass[twocolumn]{aastex631}

\shorttitle{Microlensing Detectability of Compact Multi-Planet Systems}
\shortauthors{Stephen R. Kane}

\begin{document}

\title{On the Detectability of Compact Multi-Planet Systems via
  Gravitational Microlensing}

\author[0000-0002-7084-0529]{Stephen R. Kane}
\affiliation{Department of Earth and Planetary Sciences, University of
  California, Riverside, CA 92521, USA}
\affiliation{Center for Astrobiology and Analog Worlds, University of
  California, Riverside, CA 92521, USA}
\email{skane@ucr.edu}

%%%%%%%%%%%%%%%%%%%%%%%%%%%%%%%%%%%%%%%%%%%%%%%%%%%%%%%%%%%%%%%%%%%%

\begin{abstract}

Compact multi-planet systems orbiting ultracool dwarfs, such as
TRAPPIST-1, may be a major population of terrestrial worlds with
poorly constrained demographics. Individual close-in planets are
expected to escape microlensing detection, but whether a compact
system can produce a detectable collective signal remains to be
quantified. We calculate the microlensing signal from each TRAPPIST-1
planet, and of all seven combined, if the system were a lens toward
the bulge. For a 0.09~$M_\odot$ lens at 4~kpc, $R_E \approx 1.21$~AU,
placing the planets at $s = 0.010$--$0.051~\theta_E$, 12--63 times
interior to the lensing zone.  We measure each signal as the residual
after fitting a single-lens model, which accounts for the barycentric
dipole. Because the quadrupole term cancels between the two images to
first order, the residual scales as the octupole ($q s^3$), reaching
$|\delta A / A| \sim 10^{-6}$. The seven planets combined yield
$\Delta\chi^2 \sim 10^{-3}$ even at 1~mmag precision, five orders of
magnitude below the standard detection threshold. Detection requires
increasing all planet masses by a factor of $\sim$300, whereas the
resonant chain loses its guarantee of Hill stability at a factor of
$\sim$7. Only a lens within $\sim$15~pc of the source or the Sun
places the outer planets in the lensing zone. The Nancy Grace Roman
Space Telescope is therefore not expected to detect a TRAPPIST-1-like
system through microlensing, whereas its transit channel in the same
fields can detect such planets in large numbers, marking a sharp
boundary in the Galactic census between microlensing and transit
sensitivity.

\end{abstract}

\keywords{astrobiology -- planetary systems -- gravitational lensing:
  micro -- planets and satellites: detection -- planets and satellites:
  terrestrial planets}

%%%%%%%%%%%%%%%%%%%%%%%%%%%%%%%%%%%%%%%%%%%%%%%%%%%%%%%%%%%%%%%%%%%%

\section{Introduction}
\label{intro}

Gravitational microlensing has been demonstrated as a powerful
technique for detecting exoplanets across the Galaxy, probing a region
of the planet mass and orbital separation plane that is largely
inaccessible to the transit and radial velocity methods
\citep{mao1991a,gould1992g,gaudi2012}. In the two decades since the
first microlensing planet discovery, more than 200 planets have been
found in this way, ranging from gas giants to terrestrial worlds
\citep{cassan2012,suzuki2016,zang2025}. The technique is sensitive to
cold planets beyond the snow line, to planets orbiting hosts across
the full range of stellar mass including brown dwarfs, and to
free-floating planets unbound to any host
\citep{bennett1996,sumi2011,sumi2023}. The Nancy Grace Roman Space
Telescope \citep{penny2019,johnson2020b} will conduct a Galactic Bulge
Time Domain Survey designed to detect $\sim$1,400 bound planets
through microlensing perturbations, with a mass sensitivity extending
down to Mars-mass objects. The 15-minute cadence, space-based
photometric precision ($\sigma \approx 10$~mmag at $W146 \sim 21$),
and six 72-day observing seasons together represent a significant
advance in the demographics of microlensing planets \citep{penny2019}.

A correct interpretation of these survey yields depends on whether the
resulting planet population statistics are representative of the full
Galactic census, or whether whole classes of planetary system fall
outside the method's sensitivity window. Microlensing is most
sensitive to planets near the Einstein ring radius, typically $R_E
\sim 2$--4~AU for solar-type lenses and $R_E \sim 1$--3~AU for M~dwarf
lenses \citep{griest1998,gaudi2012}. The ``lensing zone'', at
projected separations of $\sim$0.6--1.6~$R_E$, defines the region of
peak sensitivity \citep{gould1992g}, with detection efficiency falling
steeply for planets both interior and exterior to it. Simulations of the
Roman survey likewise concentrate its sensitivity on cold planets
beyond $\sim$1~AU \citep{penny2019}. \citet{park2006a} showed that
even Habitable Zone (HZ) Earth-mass planets comprise fewer than 1\% of
detectable microlensing Earths, because the HZ generally lies well
inside the lensing zone, a conclusion reinforced in the Roman era by
\citet{sajadian2021c}, who found a detection efficiency of only
$\sim$0.01\% for Earth-mass HZ planets in the Roman survey. For
ultracool dwarf (UCD) hosts, this issue is even more severe, since the
HZ is closer to the star and the entire region of close-in planetary
orbits falls deeply interior to the lensing zone. The discovery of the
TRAPPIST-1 system \citep{gillon2016,gillon2017a,luger2017b}, seven
Earth-sized planets in a compact resonant chain orbiting a
0.09~$M_\odot$ UCD at separations of 0.012--0.062~AU, presents a
planetary architecture that is highly separated from the microlensing
sensitivity peak.  Transit surveys indicate that such compact
multi-planet systems around mid- to late-M dwarfs may be common, with
occurrence rate estimates of $21^{+7}_{-5}$\% \citep{muirhead2015} to
$44^{+45}_{-33}$\% \citep{hardegreeullman2019} for planets with
periods shorter than 10~days. Thus, if these rates extend into the UCD
regime, compact terrestrial systems may constitute a substantial
fraction of all planetary systems in the Galaxy, and yet microlensing
surveys may be entirely blind to them.

Despite the importance of whether such compact systems can be
  detected by microlensing for interpreting Galactic planet
demographics, the microlensing detectability of TRAPPIST-1-like
compact architectures has received little direct
attention. \citet{shvartzvald2017b} noted that their detection of the
Earth-mass planet OGLE-2016-BLG-1195Lb around a UCD cannot set strong
constraints on the presence of planets elsewhere around UCDs at close
separations, similar to the TRAPPIST-1 system. Subsequent Keck
adaptive optics follow-up by \citet{vandorou2025a} has since revised
OGLE-2016-BLG-1195L to a $\sim$10~$M_\oplus$ sub-Neptune around a
$\sim$0.62~$M_\odot$ M~dwarf, further revealing the limits of
microlensing sensitivity to terrestrial planets around
UCDs. \citet{miyazaki2021a} explored Roman's sensitivity to
short-period planets through the xallarap effect, the orbital motion
of the source star, but found the sensitivity limited to warm Jupiters
with $M_p \gtrsim 0.5~M_{\rm Jup}$ and periods $\lesssim 30$~days, far
more massive than the TRAPPIST-1 planets. Recent work on multi-planet
microlensing with Roman \citep{fatheddin2023b,saggese2026} has
demonstrated sensitivity to two-planet systems in configurations where
both planets fall near the lensing zone, rather than the compact
short-period architectures considered here. The relative lack of
microlensing sensitivity to individual close-in planets is therefore
well understood. However, two related questions remain to be
addressed. First, a compact system places several planets within a few
percent of an Einstein radius of the host (where their perturbations
overlap near the peak of a high-magnification event), and it has not
been established whether their combined signal could be detectable
when each individual signal is not. Second, the rate at which the
observable signal vanishes interior to the lensing zone has not been
quantified for the closest orbits, and that determines the proximity
of compact systems to detection thresholds. Both of these questions
set the inner boundary of the microlensing census for a potentially
common architecture around low-mass stars.

In this paper, we quantify the microlensing detectability of a
TRAPPIST-1-like compact planetary system, with specific application to
the Roman survey. We solve the binary lens equation for each of the
seven planets as if the system were a lens toward the Galactic bulge,
measure the residual after fitting a single-lens model, and combine
the seven perturbations to evaluate the collective signal against the
Roman survey specifications. We also examine the conditions most
favorable to detection, such as higher photometric precision,
alternative lens and source distances, and a uniform expansion of the
orbits, and compare the microlensing result with the sensitivity of
Roman's transit channel. Section~\ref{formalism} presents the
microlensing formalism and establishes the relevant scales for a UCD
lens. Section~\ref{trappist} describes the TRAPPIST-1 system
parameters and their mapping into microlensing units, and
Section~\ref{roman} summarizes the Roman survey
design. Section~\ref{results} presents the binary lens calculations
and the detectability analysis, and Section~\ref{disc} discusses the
implications for planet demographics and the complementary role of
Roman's transit channel. Our conclusions are provided in
Section~\ref{conclusions}.

%%%%%%%%%%%%%%%%%%%%%%%%%%%%%%%%%%%%%%%%%%%%%%%%%%%%%%%%%%%%%%%%%%%%

\section{Microlensing Formalism}
\label{formalism}

%%%%%%%%%%%%%%%%%%%%%%%%%%%%%%%%%%%%%%%%%%%%%%%%%%%%%%%%%%%%%%%%%%%%

\subsection{Einstein Ring and Lensing Geometry}
\label{einstein}

The characteristic angular scale of a microlensing event is set by
the Einstein ring \citep{paczynski1996a}. For a lens of mass $M$ at
distance $D_L$ with a source at distance $D_S$, the angular Einstein
ring radius is
\begin{equation}
  \theta_E = \sqrt{\frac{4GM}{c^2} \frac{D_{LS}}{D_L D_S}} \,,
  \label{eq:thetaE}
\end{equation}
where $D_{LS} = D_S - D_L$. The corresponding physical Einstein
radius projected onto the lens plane is
\begin{equation}
  R_E = \theta_E D_L = 8.0727
  \sqrt{\frac{M}{M_\odot} \cdot x(1-x) \cdot \frac{D_S}{8~{\rm
        kpc}}} ~{\rm AU} \,,
  \label{eq:RE}
\end{equation}
where $x \equiv D_L / D_S$. For a standard bulge microlensing geometry
with $D_S = 8$~kpc and $D_L = 4$~kpc ($x = 0.5$), and a
TRAPPIST-1-mass host of $M = 0.09~M_\odot$, we obtain $R_E = 1.21$~AU
and $\theta_E = 0.302$~mas. This choice of $x = 0.5$ maximizes $R_E$
and therefore minimizes the projected separations $s = a/R_E$, making
it the least favorable geometry for detecting close planets. Lenses
near either distance extreme have smaller Einstein radii and
correspondingly larger $s$, so a complete treatment would weight the
detectability over the distribution of lens and source distances.
Figure~\ref{fig:separation}b and Section~\ref{demographics}
show that geometries favorable enough for detection require the
  lens to lie within $\sim$20~pc of the source or the observer and
are correspondingly rare, and so we adopt the $x = 0.5$ geometry
throughout.

The Einstein ring crossing time, which sets the characteristic
timescale of a microlensing event, is
\begin{equation}
  t_E = \frac{\theta_E}{\mu_{\rm rel}} \,,
  \label{eq:tE}
\end{equation}
where $\mu_{\rm rel}$ is the lens-source relative proper motion. For a
typical bulge event with $\mu_{\rm rel} = 5$~mas~yr$^{-1}$, we obtain
$t_E \approx 22$~days. Because $\theta_E \propto M^{1/2}$, this is
about three times shorter than for a solar-mass lens in the same
geometry.

%%%%%%%%%%%%%%%%%%%%%%%%%%%%%%%%%%%%%%%%%%%%%%%%%%%%%%%%%%%%%%%%%%%%

\subsection{Binary Lens Magnification}
\label{binarylens}

For a two-body lens consisting of a host star and a planet, the lens
equation relates the source position $\zeta$ to the image positions
$z$ in the complex plane \citep{witt1990,witt1995b}:
\begin{equation}
  \zeta = z - \frac{m_\star}{\bar{z}} - \frac{m_p}{\bar{z} - s} \,,
  \label{eq:lenseq}
\end{equation}
where $m_\star = 1/(1+q)$ and $m_p = q/(1+q)$ are the mass fractions of
the host and planet, $q = M_p / M_\star$ is the mass ratio, $s$ is the
projected star-planet separation in units of $\theta_E$, and bars
denote complex conjugation. We place the host at the origin and the
planet at position $s$ along the real axis.

Combining Equation~\ref{eq:lenseq} with its complex conjugate yields a
fifth-degree polynomial in $z$ \citep{witt1990,witt1995b}, which has
either three or five physical image solutions depending on whether the
source lies outside or inside a caustic curve. The magnification of
each image is
\begin{equation}
  \mu_j = \frac{1}{\left| \det J_j \right|} \,,
  \label{eq:mag}
\end{equation}
where the Jacobian determinant at image $z_j$ is
\begin{equation}
  \det J_j = 1 - \left| \frac{m_\star}{\bar{z}_j^2} +
  \frac{m_p}{(\bar{z}_j - s)^2} \right|^2 \,,
  \label{eq:detJ}
\end{equation}
and the total magnification is $A = \sum_j \mu_j$. Throughout the
detectability calculations of Section~\ref{results} we treat the
source as point-like. A finite source averages the magnification over
the stellar disk. In principle this averaging need not lower the
deviation along a given trajectory, since parts of the disk can sample
structure the center misses, but in the deeply finite-source regime
relevant here (Section~\ref{finitesource}), where the source spans the
caustic thousands of times over, the averaging suppresses the compact
perturbation, so the point-source deviation is an upper limit on the observable signal. The finite-source
correction is discussed separately in Section~\ref{finitesource},
where it is shown to weaken the signal further, and the numerical
treatment of an extended source is validated in
Section~\ref{validation}.

%%%%%%%%%%%%%%%%%%%%%%%%%%%%%%%%%%%%%%%%%%%%%%%%%%%%%%%%%%%%%%%%%%%%

\subsection{Central Caustic for Close Planets}
\label{caustic}

For planets with projected separations $s < 1$ (the ``close''
topology), the caustic structure consists of a central caustic near
the host star position and two small planetary caustics on the
opposite side of the host from the planet, centered near $s - 1/s$
along the binary axis \citep{chung2005,han2006a}. The central caustic
is a diamond-shaped curve whose width along the binary axis, measured
between the two on-axis cusps, is \citep{chung2005}
\begin{equation}
  \Delta\xi_\parallel \approx \frac{4q}{(s - 1/s)^2} \,,
  \label{eq:dxi}
\end{equation}
equivalently $4qs^2/(1 - s^2)^2$. For $s \ll 1$ this reduces to
$\Delta\xi_\parallel \approx 4qs^2$, so the caustic contracts rapidly
as the planet moves inward, vanishing in the limit where the planet
merges with the host. We determine the perpendicular width
$\Delta\eta_\perp$ numerically from the critical curve
(Section~\ref{results}). In the close regime it is nearly equal to
$\Delta\xi_\parallel$, so the central caustic is close to isotropic
rather than elongated. The caustic area therefore scales as $\Delta\xi
\cdot \Delta\eta \propto q^2 s^4$, making the perturbation region
vanishingly small for the mass ratios and separations characteristic
of compact terrestrial systems.

The detection of a planetary perturbation requires the source
trajectory to pass sufficiently close to the central caustic that the
deviation from a point-lens magnification exceeds the photometric
noise. For high-magnification events (impact parameter $u_0 \ll 1$),
the point-lens magnification is $A_0 \approx 1/u_0$, and the
fractional perturbation grows in proportion to $A_0$ as the trajectory
is drawn closer to the lens axis \citep{griest1998}. The amplitude is
linear in the mass ratio and, as the caustic scaling above implies,
falls with decreasing separation. We evaluate it directly from the
binary lens equation rather than from an approximate scaling relation
(Section~\ref{results}). For Earth-mass planets ($q \sim 10^{-5}$) at
$s \sim 0.01$--$0.05$ the residual after fitting a single-lens model
is of order $10^{-6}$, far below Roman's single-epoch precision of
$\sim$10~mmag for the faint main-sequence bulge sources that dominate
the microlensing sample ($W146 \approx 21$; Section~\ref{roman}). For
context, ongoing ground-based surveys such as OGLE reach a few mmag
for bright bulge sources \citep{udalski2015b}; Roman's precision
applies to faint sources and OGLE's to bright ones, but the
perturbations considered here lie below the reach of either.

%%%%%%%%%%%%%%%%%%%%%%%%%%%%%%%%%%%%%%%%%%%%%%%%%%%%%%%%%%%%%%%%%%%%

\subsection{Numerical Implementation and Validation}
\label{validation}

The fifth-degree polynomial of Section~\ref{binarylens}
\citep{witt1990} is solved by Durand-Kerner iteration, in which every root
  estimate $z_k$ is updated simultaneously as $z_k \rightarrow z_k -
  P(z_k)/\prod_{l \neq k} (z_k - z_l)$, where $P$ is the polynomial
  normalized to unit leading coefficient. The method thus refines
all five roots at once rather than by the successive deflation of
standard solvers, thereby avoiding the associated error accumulation.
Each root is substituted back into Equation~\ref{eq:lenseq} and
retained only if the residual identifies it as a physical image. The
residuals of the physical and spurious roots are separated by more
than four orders of magnitude, so this classification is unambiguous.
Critical curves are obtained by solving the corresponding quartic for
$\det J = 0$ at each phase angle and mapping the solutions into the
source plane.

Because the main result of this paper is a non-detection, we have
verified the implementation against analytic limits and against a
published planetary event, rather than relying on internal consistency
alone. In the limit $q \rightarrow 0$ the code recovers the
single-lens magnification of \citet{paczynski1986a} with a fractional
error proportional to $q$, reaching $5 \times 10^{-13}$ at $q =
10^{-12}$. The same limit is recovered as $s \rightarrow 0$, where the
two masses merge. For source positions yielding the maximum number of
images, the signed magnifications satisfy the point-mass sum rule
$\sum_j \mathrm{sgn}(\det J_j)\,\mu_j = 1$ to within $2.5 \times
10^{-10}$ across 366 configurations spanning $s = 0.6$--2.0 and $q =
10^{-3}$--1. The number of images is everywhere three or five, the
magnification is symmetric about the star--planet axis to one part in
$10^{12}$, and the caustic topology changes from a single curve to two
at the analytic wide-binary boundary $s_w^2 = (1 + q^{1/3})^3/(1+q)$
\citep{dominik1999b}
for every mass ratio tested. The numerically determined central
caustic width reproduces the analytic expression of \citet{chung2005},
$\Delta\xi = 4q/(s - 1/s)^2$, to five significant figures on both
sides of $s = 1$.

As an end-to-end test we reproduced the planetary event
OGLE-2005-BLG-390 \citep{beaulieu2006}, adopting only the published
model parameters $s = 1.610$, $q = 7.6 \times 10^{-5}$, $u_0 = 0.359$,
$t_E = 11.03$~d, $\rho_\star = 0.0256$, and trajectory angle
$2.756$~rad.  The computed peak magnification of $2.92$ matches the
published value of $\approx$2.8, the planetary caustic center falls at
$s - 1/s = 0.9889$, and the anomaly occurs $10.15$~days from peak,
against the observed interval of approximately ten days between the
peak on 2005 July 31 and the deviation on 2005 August 10. Integrating
over a uniform source of the published angular size yields a peak
deviation of $18.5$\%, consistent with the reported $\sim$15\% flux
excursion once blending is accounted for. This event is strongly
finite-source dominated whereby the source radius exceeds the
planetary caustic dimensions, so the extended source encompasses a
caustic that the source center never crosses.

%%%%%%%%%%%%%%%%%%%%%%%%%%%%%%%%%%%%%%%%%%%%%%%%%%%%%%%%%%%%%%%%%%%%

\section{The TRAPPIST-1 System}
\label{trappist}

The TRAPPIST-1 system consists of seven terrestrial planets orbiting a
$0.0898 \pm 0.0023~M_\odot$ UCD star \citep{gillon2017a,agol2021} at a
distance of 12.47~pc \citep{vallenari2023}. The system is among the
most precisely characterized multi-planet systems, with transit timing
variations providing mass determinations at the few-percent level
\citep{grimm2018,agol2021}. Table~\ref{tab:params} summarizes the
planetary parameters relevant to this study.

\begin{deluxetable}{lccccc}
  \tablecaption{TRAPPIST-1 planet parameters and microlensing
    quantities for a 0.09~$M_\odot$ lens at $D_L = 4$~kpc toward a
    bulge source at $D_S = 8$~kpc ($R_E =
    1.21$~AU).\label{tab:params}}
  \tablehead{
    \colhead{Planet} &
    \colhead{$a$ (AU)} &
    \colhead{$M_p$ ($M_\oplus$)} &
    \colhead{$s$} &
    \colhead{$q$} &
    \colhead{$\Delta\xi_\parallel$}
  }
  \startdata
  b & 0.01154 & 1.374 & 0.0095 & $4.60 \times 10^{-5}$ & $1.67 \times 10^{-8}$ \\
  c & 0.01580 & 1.308 & 0.0131 & $4.37 \times 10^{-5}$ & $2.99 \times 10^{-8}$ \\
  d & 0.02228 & 0.388 & 0.0184 & $1.30 \times 10^{-5}$ & $1.76 \times 10^{-8}$ \\
  e & 0.02928 & 0.692 & 0.0242 & $2.31 \times 10^{-5}$ & $5.43 \times 10^{-8}$ \\
  f & 0.03853 & 1.039 & 0.0319 & $3.48 \times 10^{-5}$ & $1.41 \times 10^{-7}$ \\
  g & 0.04688 & 1.321 & 0.0388 & $4.42 \times 10^{-5}$ & $2.66 \times 10^{-7}$ \\
  h & 0.06193 & 0.326 & 0.0512 & $1.09 \times 10^{-5}$ & $1.15 \times 10^{-7}$ \\
  \enddata
  \tablecomments{Orbital and mass parameters from
    \citet{agol2021}. The projected separation $s$ and mass ratio $q$
    are computed assuming the orbital semi-major axis equals the
    projected separation. The central caustic width
    $\Delta\xi_\parallel$, in units of $\theta_E$, is determined
    numerically from the critical curve.}
\end{deluxetable}

An interesting feature of Table~\ref{tab:params} is the extremely
small projected separations. All seven planets fall at $s =
0.0095$--$0.051$, placing them 12--63 times closer to the host than
the inner edge of the conventional lensing zone at $s = 0.6$
\citep{gould1992g}. The mass ratios of $q \sim 10^{-5}$ are also near
the floor of current microlensing sensitivity. Ground-based surveys
detect planets most efficiently at $q \gtrsim 10^{-4}$
\citep{suzuki2016}, and only recently have samples reached $q \sim
10^{-5}$ \citep{zang2025}. The combination of small $s$ and small $q$
places the TRAPPIST-1 planets in a region of parameter space that is observationally challenging, and in which survey detection
  efficiency calculations find negligible sensitivity
  \citep{suzuki2016,penny2019}.

%%%%%%%%%%%%%%%%%%%%%%%%%%%%%%%%%%%%%%%%%%%%%%%%%%%%%%%%%%%%%%%%%%%%

\section{Roman Survey Parameters}
\label{roman}

The Roman Galactic Bulge Time Domain Survey will observe $\sim$2
deg$^2$ toward the Galactic center with the following specifications
\citep{penny2019,johnson2020b}: a cadence of 15 minutes in the wide
$W146$ filter (1--2~$\mu$m), photometric precision of $\sigma \approx
0.01$~mag at $W146 \approx 21$, and six 72-day observing seasons
spread across a 5-year mission. The survey is expected to detect
$\sim$27,000 microlensing events with $u_0 < 1$ and $\sim$1,400 bound
planetary perturbations \citep{penny2019}.

For the detection of a planetary perturbation, we adopt the standard
criterion of total $\Delta\chi^2 > 160$
\citep{gould1992g,gaudi2012}, where
\begin{equation}
  \Delta\chi^2 = \sum_i \left( \frac{A_{\rm binary}(t_i) - A_{\rm
      1L1S}(t_i)}{\sigma_i \, A_{\rm 1L1S}(t_i)} \right)^2 \,,
  \label{eq:dchi2}
\end{equation}
where $A_{\rm 1L1S}$ is the best-fitting single-lens model, the sum
runs over data points sampled at Roman's cadence during the event, and
$\sigma_i$ is the fractional flux uncertainty. Numerator and
denominator are both magnifications, so the ratio is dimensionless. We
adopt a uniform $\sigma_i = 0.01$, the fractional precision equivalent
to $\sim$10~mmag for main-sequence source stars near $W146 \approx
21$. A uniform fractional error is a simplified stand-in for a full
photon-based noise model. During a high-magnification event, the
source brightens by factors of hundreds, so the true per-epoch
precision near peak is better than this baseline, which would lower
$\sigma_i$ and raise every reported $\Delta\chi^2$. Our fixed 1\%
therefore understates rather than overstates detectability. Even so,
the resulting significances fall short of the detection criterion by
many orders of magnitude, so no realistic photometric model closes the
gap, and adopting one would only reinforce the non-detection at the
true planet masses.

%%%%%%%%%%%%%%%%%%%%%%%%%%%%%%%%%%%%%%%%%%%%%%%%%%%%%%%%%%%%%%%%%%%%

\section{Results}
\label{results}

%%%%%%%%%%%%%%%%%%%%%%%%%%%%%%%%%%%%%%%%%%%%%%%%%%%%%%%%%%%%%%%%%%%%

\subsection{Perturbation Amplitudes}
\label{amplitudes}

We solve the binary lens equation (Equation~\ref{eq:lenseq})
numerically for each TRAPPIST-1 planet, computing the total
magnification along source trajectories parameterized by impact
parameter $u_0$ and trajectory angle $\phi$ relative to the binary
axis. The observable planetary signal is not the raw difference
between the binary light curve and a point lens fixed at the host
star. For a close companion, the images form near the Einstein ring of
the total mass, and to leading order the system acts as a single lens
located at its center of mass, displaced from the host by $\approx q
s$ in units of $\theta_E$ \citep{dominik1999b}. That displacement is
not an observable, rather it is absorbed when the standard single-lens
parameters $(t_0, u_0, t_E)$ are fitted to the light curve, exactly as
in the analysis of a real event. We therefore define the residual
deviation as
\begin{equation}
  \frac{\delta A}{A_0} = \frac{A_{\rm binary} -
  A_{\rm 1L1S}}{A_{\rm 1L1S}} \,,
  \label{eq:fracdev}
\end{equation}
where $A_{\rm 1L1S}$ is the best-fitting single-lens model
\citep{paczynski1986a}. For the individual-planet residuals of
Table~\ref{tab:perturbations} we fit $t_0$, which absorbs the
barycentric displacement. Also refitting $u_0$ and $t_E$ changes the
residual by only a few percent (Section~\ref{scaling}). For the
combined signal of all seven planets, all three parameters are fitted
(Section~\ref{combined}). This removes the unobservable dipole and
leaves only the genuine, higher-order distortion that a single lens
cannot reproduce.

\begin{figure*}
  \includegraphics[angle=270,width=0.95\linewidth]{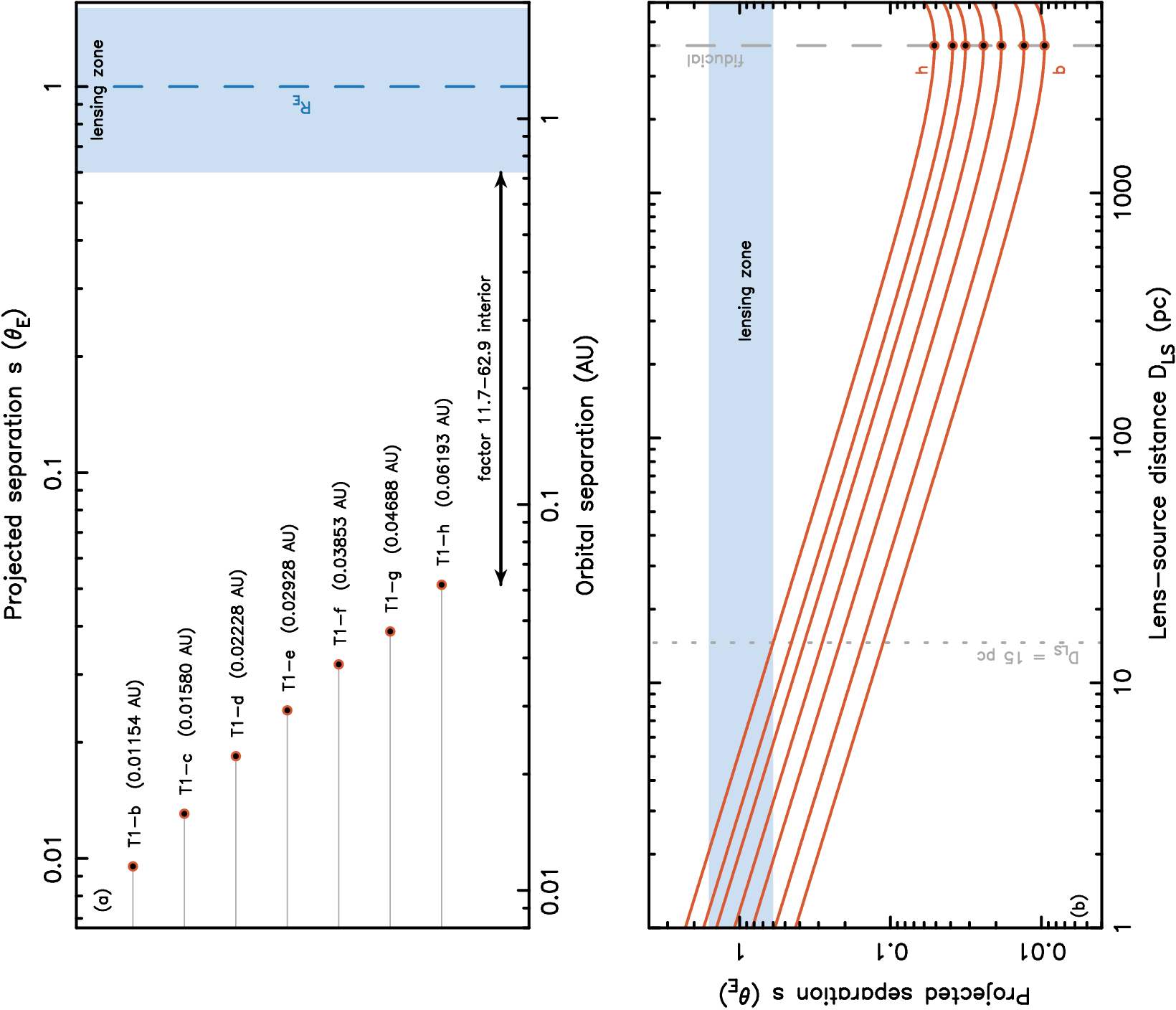}
  \caption{(a) Orbital separations of the seven TRAPPIST-1 planets
    (orange points) compared to the lensing zone (0.6--1.6~$R_E$, blue
    shaded region) for a 0.09~$M_\odot$ lens at $D_L = 4$~kpc toward a
    Galactic bulge source at $D_S = 8$~kpc. The dashed line marks the
    Einstein ring radius $R_E = 1.21$~AU. All seven planets fall
    12--63 times interior to the inner edge of the lensing zone,
    placing them in the close-planet regime where microlensing
    perturbations are minimal. (b) Projected separation of each
    planet, from b (lowest) to h (highest), as the lens is moved
    toward the source at fixed $D_S = 8$~kpc, with the lensing zone
    shaded.  The dashed line marks the fiducial geometry of panel (a),
    where $R_E$ is largest and every separation smallest, so the
    curves rise on either side of it. The dotted line marks $D_{LS} =
    15$~pc, inside which the outermost planet enters the lensing
    zone. The same curves apply with $D_L$ in place of $D_{LS}$ for a
    lens near the observer.}
  \label{fig:separation}
\end{figure*}

Figure~\ref{fig:separation} illustrates the fundamental geometric
mismatch between the TRAPPIST-1 planet separations and the
microlensing sensitivity zone. The planets span 0.012--0.062~AU, while
the lensing zone covers 0.73--1.94~AU for a 0.09~$M_\odot$ lens,
equivalent to a factor of 12--63 gap in separation space.

Table~\ref{tab:perturbations} presents the maximum residual deviation
$|\delta A / A_0|_{\rm max}$ for each planet at impact parameters $u_0
= 0.001$, 0.005, and 0.01, corresponding to peak magnifications of $A
\approx 1000$, 200, and 100 for a point source, respectively (see
Section~\ref{finitesource}). The source trajectory is aligned with the
star--planet axis ($\phi = 0$). After the single-lens model is fitted,
the largest single-planet residual is that of planet~g ($s =
  0.039$, $q = 4.42 \times 10^{-5}$) at $u_0 = 0.001$, $6.6 \times
10^{-6}$, or $0.007$~mmag, and the smallest is that of planet~b,
  at $1.1 \times 10^{-7}$. The deviations scale approximately linearly
with $1/u_0$, so lower magnifications yield proportionally weaker
signals.

\begin{deluxetable}{lcccc}
  \tablecaption{Maximum fractional magnification deviation for each
    TRAPPIST-1 planet at selected impact parameters, measured as the
    residual after fitting a single-lens model (in which only the
      time of peak, $t_0$, is fitted) to the
    binary light curve, so that the unobservable
    center-of-mass displacement is removed.\label{tab:perturbations}}
  \tablehead{
    \colhead{Planet} &
    \colhead{$|\delta A/A|_{\rm max}$} &
    \colhead{$|\delta A/A|_{\rm max}$} &
    \colhead{$|\delta A/A|_{\rm max}$} &
    \colhead{Fraction of} \\
    \colhead{} &
    \colhead{$u_0 = 0.001$} &
    \colhead{$u_0 = 0.005$} &
    \colhead{$u_0 = 0.01$} &
    \colhead{$\sigma_{\rm Roman}$}
  }
  \startdata
  b & $1.06 \times 10^{-7}$ & $2.44 \times 10^{-8}$ & $1.44 \times 10^{-8}$ & 0.001\% \\
  c & $2.56 \times 10^{-7}$ & $5.69 \times 10^{-8}$ & $3.22 \times 10^{-8}$ & 0.003\% \\
  d & $2.12 \times 10^{-7}$ & $4.56 \times 10^{-8}$ & $2.49 \times 10^{-8}$ & 0.002\% \\
  e & $8.52 \times 10^{-7}$ & $1.80 \times 10^{-7}$ & $9.66 \times 10^{-8}$ & 0.009\% \\
  f & $2.90 \times 10^{-6}$ & $6.06 \times 10^{-7}$ & $3.20 \times 10^{-7}$ & 0.029\% \\
  g & $6.63 \times 10^{-6}$ & $1.37 \times 10^{-6}$ & $7.18 \times 10^{-7}$ & 0.066\% \\
  h & $3.76 \times 10^{-6}$ & $7.72 \times 10^{-7}$ & $3.99 \times 10^{-7}$ & 0.038\% \\
  \enddata
  \tablecomments{The final column gives the best-case ($u_0 = 0.001$)
    perturbation as a fraction of Roman's $1\sigma$ photometric
    precision (10~mmag). All entries assume a point source, which is
      optimistic at $u_0 = 0.001$, where the magnification of a
      main-sequence bulge source saturates near 1040 and the
      finite-source residual is smaller
      (Section~\ref{finitesource}).}
\end{deluxetable}

%%%%%%%%%%%%%%%%%%%%%%%%%%%%%%%%%%%%%%%%%%%%%%%%%%%%%%%%%%%%%%%%%%%%

\subsection{Light Curve Analysis}
\label{lightcurves}

\begin{figure*}
  \includegraphics[angle=270,width=\linewidth]{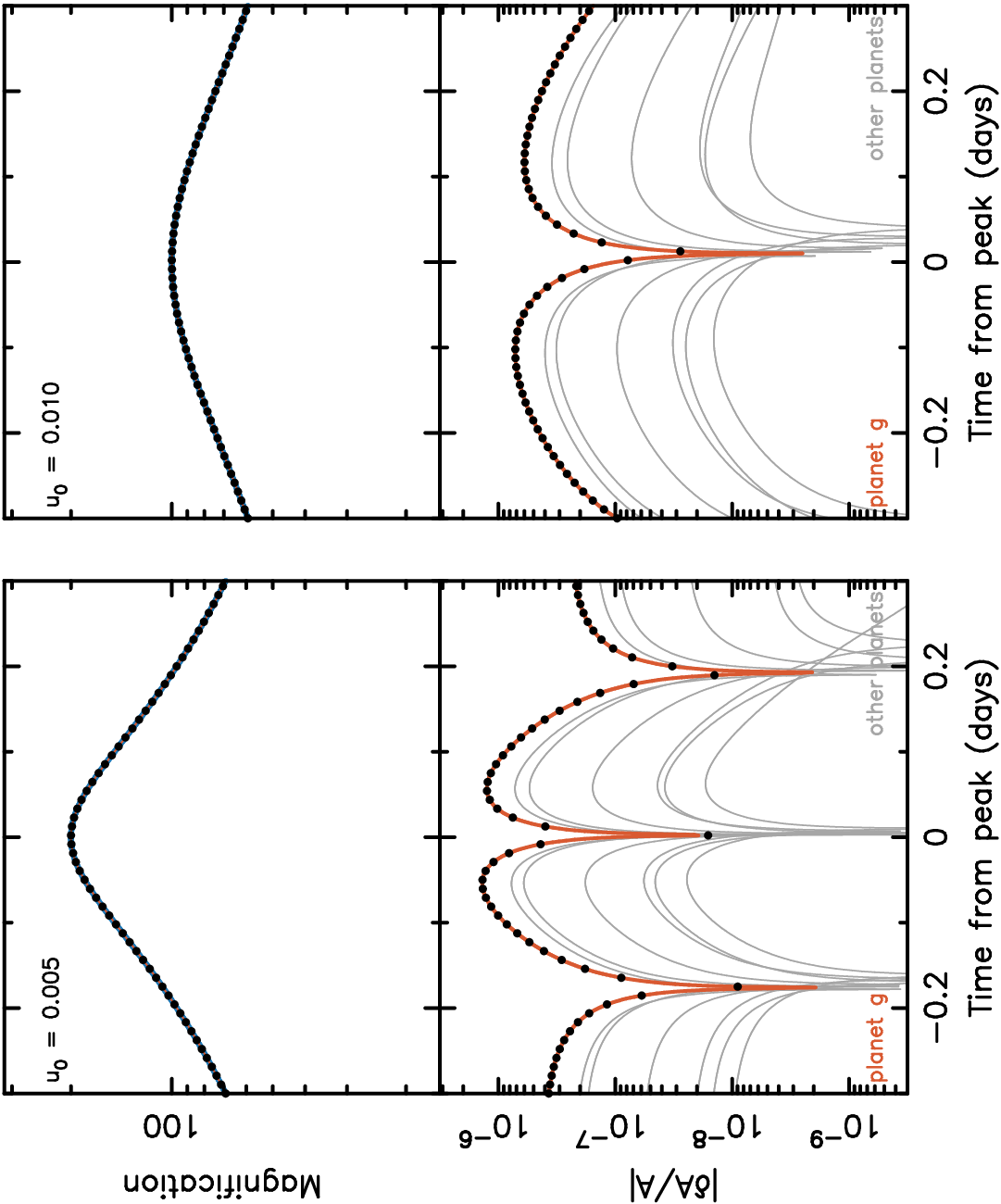}
  \caption{Light curves and fractional deviations for the TRAPPIST-1
    planets acting as microlenses, at impact parameters $u_0 = 0.005$
    (left) and $u_0 = 0.01$ (right). Upper panels: the single-lens
    magnification (thick blue) with the binary lens curve for planet~g
    drawn over it (thin orange). Lower panels: the fractional
    deviation $|\delta A / A_0|$ for all seven planets, with g in
    orange and the remaining six in gray. Filled circles in both rows
    mark Roman's 15-minute sampling. Roman's $1\sigma$ threshold of
    10~mmag lies far above the plotted range and is omitted from the
    lower panels.}
  \label{fig:lightcurves}
\end{figure*}

Figure~\ref{fig:lightcurves} shows representative light curves at two
impact parameters ($u_0 = 0.005$ and $0.01$). The upper panels compare
the binary lens magnification with that of a single lens of the same
total mass; the two are indistinguishable at Roman's photometric
precision. The lower panels show the fractional deviations for all
seven planets. The perturbation tracks the magnification profile
across the entire event, rather than being confined to a narrow
interval around peak magnification, with a full width at half maximum
of $\approx$300--330~minutes at $u_0 = 0.005$. Roman's 15-minute
cadence therefore places about twenty exposures across the
perturbation itself, such that sampling is not the limitation. After
the single-lens model is fitted, the largest residual at $u_0 = 0.005$
(planet~g) peaks at $1.4 \times 10^{-6}$. Each deviation curve (Figure~\ref{fig:lightcurves},
lower panels) shows several lobes separated by narrow downward
spikes. These arise because the plotted quantity is the residual after
a single-lens model is fitted to the light curve. The fit removes the
smooth dipole component, leaving a higher-order residual that changes
sign several times across the event, and each sign change appears as a
sharp dip of the modulus on a logarithmic scale.

%%%%%%%%%%%%%%%%%%%%%%%%%%%%%%%%%%%%%%%%%%%%%%%%%%%%%%%%%%%%%%%%%%%%

\subsection{Total Detection Significance}
\label{dchi2}

Even if the instantaneous perturbation is below the single-epoch
precision, a sufficiently long perturbation sampled at multiple epochs
could in principle accumulate a detectable $\Delta\chi^2$. The source
never enters the central caustic for any of the TRAPPIST-1 planets, so
the deviation is not a caustic crossing but a smooth perturbation of
the magnification profile near peak. Its full width at half maximum at
$u_0 = 0.001$ is $\Delta t_{\rm pert} \approx 65~{\rm min}$, so
Roman's 15-minute cadence places roughly four exposures above half
maximum.  Summing over a full Einstein
crossing at Roman cadence, the individual planets give $\Delta\chi^2$
between $\sim$$5 \times 10^{-10}$ and $\sim$$2 \times 10^{-6}$, and
the seven combined give $\sim$$1 \times 10^{-5}$
(Section~\ref{combined}).  Because the
perturbation spans several exposures, the sum depends only weakly on
the arbitrary offset between the cadence grid and the peak.  The
implied detectability is negligible across all impact parameters and
trajectory angles considered.

%%%%%%%%%%%%%%%%%%%%%%%%%%%%%%%%%%%%%%%%%%%%%%%%%%%%%%%%%%%%%%%%%%%%

\subsection{Finite Source Effects}
\label{finitesource}

An additional suppression of the perturbation arises from the finite
angular size of the source star \citep{witt1994b,gould1994b}. The
normalized source radius is
\begin{equation}
  \rho_\star = \frac{\theta_\star}{\theta_E} = x \,
  \frac{R_\star}{R_E} \,,
  \label{eq:rho}
\end{equation}
where $R_\star$ is the source star radius, and the factor $x$ arises
because $\theta_\star$ is set by $D_S$ while $\theta_E$ is set by
$D_L$. For a main-sequence source ($R_\star = 1~R_\odot$), $\rho_\star
= 0.0019$; for a K-giant source ($R_\star = 10~R_\odot$), $\rho_\star
= 0.019$. The central caustic widths for all TRAPPIST-1 planets
($\Delta\xi_\parallel \sim 2 \times 10^{-8}$--$3 \times 10^{-7}$;
Table~\ref{tab:params}) are four to five orders of magnitude smaller
than $\rho_\star$ for a main-sequence source, and smaller still
relative to a giant. The caustic is thus decisively unresolved and the
source disk spans it thousands of times over, so the finite-source
magnification averages over the caustic entirely. In this deeply
finite-source regime, the averaging suppresses the compact residual
rather than enhancing it, unlike the marginally resolved case where
part of the disk can sample a caustic the center misses. Effects of
this kind are routinely measured in observed events
\citep[e.g.,][]{yoo2004b}. The point-source residuals of Table~\ref{tab:perturbations} are
therefore upper limits on the true finite-source signal.

This averaging is most pronounced at the smallest impact parameter. A
main-sequence source has $\rho_\star = 0.0019$, so at $u_0 = 0.001$ the lens lies within the projected stellar
  disk near peak, and the
uniform-disk magnification saturates at $\sqrt{1 + 4/\rho_\star^2}
\approx 1040$ \citep{witt1994b} rather than diverging. While the lens lies within the projected disk, the disk averages
  the residual over position angles about the lens at which the
  octupole term changes sign, so that it largely cancels. We have
  verified this for planet~g by computing the uniform-disk
  magnification of the binary lens through contour integration along
  the image boundaries, and fitting a finite-source single-lens model
  in $t_0$. The peak residual falls from $6.6 \times 10^{-6}$ to $2
  \times 10^{-6}$, reached at the limb crossings, and the summed
  squared residual is smaller by a factor of eight. At $u_0 = 0.005$ and $0.01$, the trajectory
lies outside the disk and the point-source treatment is accurate to
better than a few percent. In every case considered here, the finite
size of the source lowers the already negligible residual.

%%%%%%%%%%%%%%%%%%%%%%%%%%%%%%%%%%%%%%%%%%%%%%%%%%%%%%%%%%%%%%%%%%%%

\subsection{Combined Perturbation from All Seven Planets}
\label{combined}

A compact system is an eight-body lens, not seven independent two-body
lenses, and its combined signal cannot be found by summing the
individual host-centered perturbations: the displacement of each
planet is measured relative to the common system barycenter, about
which the net dipole vanishes by construction. We therefore compute
the magnification of the full eight-body lens (the host and all seven
planets) at each epoch. Away from the lens masses the lens equation
has a major and a minor image, which we find by Newton iteration
started from the corresponding point-lens image positions, accepting
each only if it has the correct parity and the two are distinct. The additional images near the lens masses are strongly demagnified and
are omitted. For a single planet this calculation reproduces the exact
binary-lens magnification of Section~\ref{validation} to
$\sim$$10^{-10}$. We then fit a single-lens model to the full light
curve, so that the net barycentric dipole is removed in one fit.
Linear superposition of the seven binary-lens perturbations,
  correct only to first order in the planet masses, agrees with this
  to about 20\% in $\Delta\chi^2$ at the true masses.

Placing all seven planets on the same side of the axis maximizes the
coherent addition and yields a combined residual of $1.5 \times
10^{-5}$ at $u_0 = 0.001$. Realistic configurations, with the planets
at the various orbital phases they occupy during an event, are
expected to give smaller values, since the higher-order residuals of
the individual planets then no longer add in phase.  Summing at Roman cadence over
the $\pm 0.06\,t_E$ ($\pm$1.3-day) window around the peak ($\sim$250
epochs), which captures all but a few percent of the signal from a
full Einstein crossing, the total
\begin{equation}
  \Delta\chi^2 = \sum_i \left( \frac{A_{\rm total}(t_i) - A_{\rm
      1L1S}(t_i)}{\sigma_i \, A_{\rm 1L1S}(t_i)} \right)^2 \approx 9.5
  \times 10^{-6}
  \label{eq:combined}
\end{equation}
for the maximally coherent configuration, more than seven orders of
magnitude short of the detection criterion of $\Delta\chi^2 = 160$.
At 1~mmag precision the same signal gives $\Delta\chi^2 \approx 9.5
\times 10^{-4}$, still more than five orders of magnitude short. Moreover, the
  threshold of 160 applies to the comparison of single-lens and
  binary-lens fits, which differ by three parameters. A model of the
  full system adds at least $3 \times 7 = 21$ parameters, so the same
  $\Delta\chi^2$ would carry less significance, and finding the
  correct solution in such a parameter space would itself be a major
  challenge.

%%%%%%%%%%%%%%%%%%%%%%%%%%%%%%%%%%%%%%%%%%%%%%%%%%%%%%%%%%%%%%%%%%%%

\subsection{Sensitivity Map}
\label{sensitivity}

\begin{figure*}
  \includegraphics[angle=270,width=\linewidth]{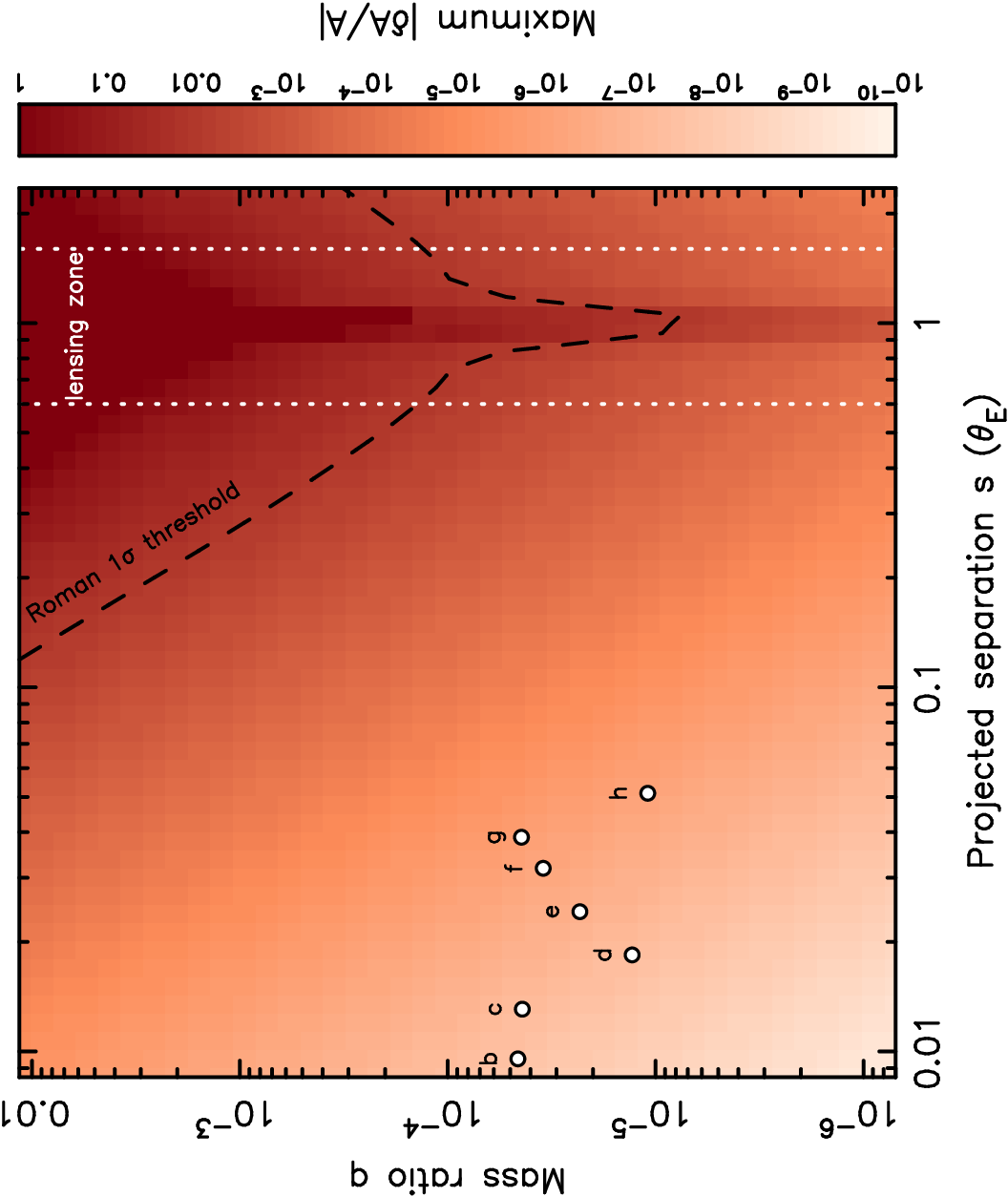}
  \caption{Two-dimensional map of the maximum microlensing
    perturbation amplitude in the projected separation ($s$) and mass
    ratio ($q$) plane, evaluated at $u_0 = 0.005$. The color scale
    shows $|\delta A / A_0|_{\rm max}$, with the Roman $1\sigma$
    single-epoch precision (10~mmag) marked by the black dashed
    contour. The seven TRAPPIST-1 planets, lettered b--h, all fall in
    the region of negligible perturbation, four to six orders of
    magnitude below that precision. White dotted lines bound the
    conventional lensing zone ($s = 0.6$--$1.6$), within which the
    resonant caustic at $s \approx 1$ produces deviations exceeding
    unity. The color scale saturates there, since any deviation beyond
    100\% is completely detectable.}
  \label{fig:sensitivity}
\end{figure*}

To place the TRAPPIST-1 results in the broader context of microlensing
planet sensitivity, Figure~\ref{fig:sensitivity} presents a
two-dimensional map of the maximum perturbation amplitude in the $(s,
q)$ parameter space. The color scale shows $|\delta A / A_0|_{\rm
  max}$ computed from the binary lens equation at $u_0 = 0.005$, with
the Roman $1\sigma$ single-epoch precision indicated by the black
dashed contour. 

The sensitivity map confirms that the microlensing blind spot for
compact systems is not specific to the TRAPPIST-1 architecture but
extends across a broad swath of parameter space at $s \lesssim 0.2$
and $q \lesssim 10^{-4}$. For a UCD lens, $s = 0.2$ corresponds to an
orbital period of about 145~days, so any compact system with periods
shorter than roughly five months would fall in this insensitive
region.

%%%%%%%%%%%%%%%%%%%%%%%%%%%%%%%%%%%%%%%%%%%%%%%%%%%%%%%%%%%%%%%%%%%%

\section{Discussion}
\label{disc}

%%%%%%%%%%%%%%%%%%%%%%%%%%%%%%%%%%%%%%%%%%%%%%%%%%%%%%%%%%%%%%%%%%%%

\subsection{Implications for Galactic Planet Demographics}
\label{demographics}

The results presented here establish that Roman is not expected to
detect a TRAPPIST-1-like compact system through its microlensing
channel. Because no individual such system would produce a detectable
signal, a microlensing non-detection places no useful constraint on
the Galactic occurrence rate of these systems. The absence of such
systems in microlensing data is entirely consistent with occurrence
rates of 20--44\% as estimated from transit surveys
\citep{muirhead2015,hardegreeullman2019}, and equally consistent with
much higher or lower rates.

The statistical opportunity for detection is also severely limited.
Of Roman's $\sim$27,000 expected microlensing events, we adopt a
fiducial 5\% as the fraction with UCD lenses ($M \lesssim
0.1~M_\odot$), a figure consistent with the low-mass end of the
stellar mass function in Galactic models, such as that of
\citet{robin2003}, though the precise value depends on the adopted
mass function and on the mass dependence of the event rate. Of these,
a fraction equal to $u_0$, or 1\%, will be high-magnification events
with $u_0 < 0.01$, since impact parameters are distributed uniformly
in $u_0$. This is the regime where the perturbations are
largest. Together, these yield $\sim$$27{,}000 \times 0.05 \times 0.01
\approx 14$ high-magnification UCD events over the full Roman
mission. Even with a TRAPPIST-1-like occurrence rate of 20\%, this
provides only $\sim$3 opportunities where the system would be present
as a lens, and our analysis shows that none of these would produce a
detectable signal.

Lens geometries more favorable than the fiducial case are also
rare. At fixed source distance, moving the lens toward the source
shrinks $R_E$ and raises every projected separation
(Figure~\ref{fig:separation}), but the outermost planet, h, reaches
the inner edge of the lensing zone only when the lens lies within
$\sim$15~pc of the source, and planet g only within $\sim$8~pc. The
angular Einstein radius is then $\sim$20 times smaller than in the
fiducial case, so a main-sequence source has $\rho_\star \approx
0.04$, and the event lasts only about a day for $\mu_{\rm rel} =
5$~mas~yr$^{-1}$. To estimate how often such geometries occur, we
weight lens-source pairs by their Einstein radius using the G2 bar
model of \citet{dwek1995}, with the scale lengths and $20\degr$
orientation adopted by \citet{han2003b}. For sightlines with $|l|
\lesssim 2\degr$ and $b$ between $-1\degr$ and $-4\degr$, planet h
reaches the lensing zone in only (3--6)$\times 10^{-4}$ of bulge
self-lensing events, a result insensitive to the bar orientation and
to the weighting of source distances. Disk lenses add events with
larger Einstein radii, so this is an upper limit on the fraction of
all events. Planets in the lensing zone produce anomalies from either the
  central or the planetary caustics, and so can be detected outside
  high-magnification events, and the relevant sample is all
$\sim$1,350 UCD events. With an occurrence rate of
20--44\%, Roman is therefore expected to observe only $\sim$0.1--0.4
TRAPPIST-1-like systems in which any planet lies within the lensing
zone, and a detection would further require the source to pass close to one of the caustics while its large angular size smears the
perturbation. The mirror case of a lens within $\sim$15~pc of the
observer is rarer still. The 10~pc census of \citet{reyle2021} implies
a local density of $\approx$0.11 stars and brown dwarfs per pc$^3$, so
fewer than 0.1 objects of any kind are expected within 15~pc in
Roman's $\sim$2~deg$^2$ fields.

Intermediate lens distances do not alter this conclusion. Moving
  the lens toward either the source or the observer raises every
  projected separation by the factor $R_E(x = 0.5)/R_E$, so we
  recomputed the combined $\Delta\chi^2$ of the full eight-body lens
  at 1~mmag precision and $u_0 = 0.001$ over a range of lens
  distances, scaling $t_E$ with $\theta_E$. Doubling the separations
  requires the lens to lie $\approx$0.54~kpc from the source or the
  observer. Near the source, $t_E$ falls to $\sim$6~days and
  $\Delta\chi^2 \approx 3 \times 10^{-3}$, before the suppression
  from $\rho_\star \approx 0.007$. Near the observer, $\theta_E$ rises
  to 1.1~mas and, for a typical transverse velocity of
  30~km~s$^{-1}$, $t_E \approx 35$~days and $\Delta\chi^2 \approx
  0.1$. Closer to the source, $t_E$ falls below a day and the peak is
  barely sampled, so $\Delta\chi^2$ remains below 0.1 down to $D_{LS}
  \approx 12$~pc, where the outer planets already lie in the lensing
  zone. Close to the observer, $\Delta\chi^2$ reaches 160 only for
  $D_L \lesssim 20$~pc. Counting every star and brown dwarf at the
  local density of \citet{reyle2021}, lenses within that distance in
  Roman's fields are expected to produce only $\sim$0.006 events with
  $u_0 < 1$ over the survey, and $\sim$$10^{-5}$ with $u_0 \lesssim
  10^{-3}$.

The demographic implications extend beyond UCD hosts, since for any
host star mass, planets with orbital periods shorter than a few days
will have projected separations well inside the lensing zone. The
microlensing planet population is therefore systematically incomplete
at short periods, complementing the transit method's incompleteness at
long periods \citep{gaudi2012,winn2015}. A complete picture of the
Galactic planet mass function across all orbital separations will
require combining microlensing statistics (sensitive at
$\sim$0.5--10~AU) with transit statistics (sensitive at
$\lesssim$0.5~AU) from surveys of the same Galactic fields, an
approach already demonstrated through a combination of radial
velocity, microlensing, and imaging data
\citep{clanton2014a,clanton2016,zhu2021}.

%%%%%%%%%%%%%%%%%%%%%%%%%%%%%%%%%%%%%%%%%%%%%%%%%%%%%%%%%%%%%%%%%%%%

\subsection{The Central Caustic and Scaling Relations}
\label{scaling}

After the single-lens model is fitted, the residual deviation in the
close-planet regime is linear in the mass ratio and, empirically,
cubic in the projected separation. Over $s = 0.01$--0.1 and $q =
10^{-5}$--$3 \times 10^{-4}$ at $u_0 = 0.001$ we measure
\begin{equation}
  \left. \frac{\delta A}{A_0} \right|_{\rm max} \approx 2.6 \times
  10^{3} \, q \, s^{3} \,,
  \label{eq:pertfit}
\end{equation}
with the coefficient constant across that range to better than 1\%.
The linear dependence on $q$ is expected since the perturbing mass
enters the lens equation linearly. The cubic dependence on $s$ follows
from the close-binary multipole expansion \citep{dominik1999b}. About
the barycenter, the dipole vanishes, and the leading correction to a
single lens is the quadrupole, of order $q s^2$, which also sets the
size of the central caustic \citep{chung2005}. However, near the peak
of a high-magnification event, the two images lie close to the
Einstein ring at nearly opposite position angles and have Jacobian
determinants of opposite sign. Treating the planet as a weak
perturbation to the point lens \citep{an2005}, a perturbing multipole
of even order changes both determinants with the same sign, so its
first-order contributions to the two image magnifications cancel in
the total, and the quadrupole enters only at second order, as $(q
s^2/u)^2$. A multipole of odd order changes them with opposite signs
and survives at first order. The leading observable term is therefore
the octupole, of order $q s^3/u$ which, for a trajectory parallel to
the binary axis, is antisymmetric in time about the peak. The
calculations performed in this work confirm this behavior. Relative to
a single lens placed at the barycenter, the residual scales as $q s^3$
before any fitting, is linear in $q$, and is dominated by its
time-antisymmetric part, while refitting $(t_0, u_0, t_E)$ reduces it
by only a few percent. The fit is essential mainly because it locates
the barycenter, removing the dipole that a host-centered single-lens
model would leave. The perturbation therefore falls faster than the
caustic shrinks as the planet moves inward, vanishing as $s
\rightarrow 0$ where the planet merges with the host and the system
reverts to a single lens. Thus, the primary reason compact systems are
inaccessible is that compactness itself suppresses the observable
signal.

For the best case (TRAPPIST-1g), Equation~\ref{eq:pertfit} yields $6.7
\times 10^{-6}$, within 5\% of the fitted residual of $6.6 \times
10^{-6}$ in Table~\ref{tab:perturbations}. Extrapolating
Equation~\ref{eq:pertfit} well beyond its fitted range, reaching the
Roman single-epoch precision of $\delta A / A_0 = 0.01$ with an
Earth-mass planet ($q \sim 3 \times 10^{-5}$) requires $s \approx
0.5$, corresponding to a semi-major axis of $\approx$0.61~AU for a UCD
lens. That separation lies just short of the inner edge of the
conventional lensing zone ($s = 0.6$) and is roughly ten times wider
than the orbit of TRAPPIST-1h. An Earth-mass planet orbiting a UCD
therefore becomes detectable only once it has left the compact regime
altogether and approached the lensing zone.

\begin{figure*}
  \includegraphics[angle=270,width=\linewidth]{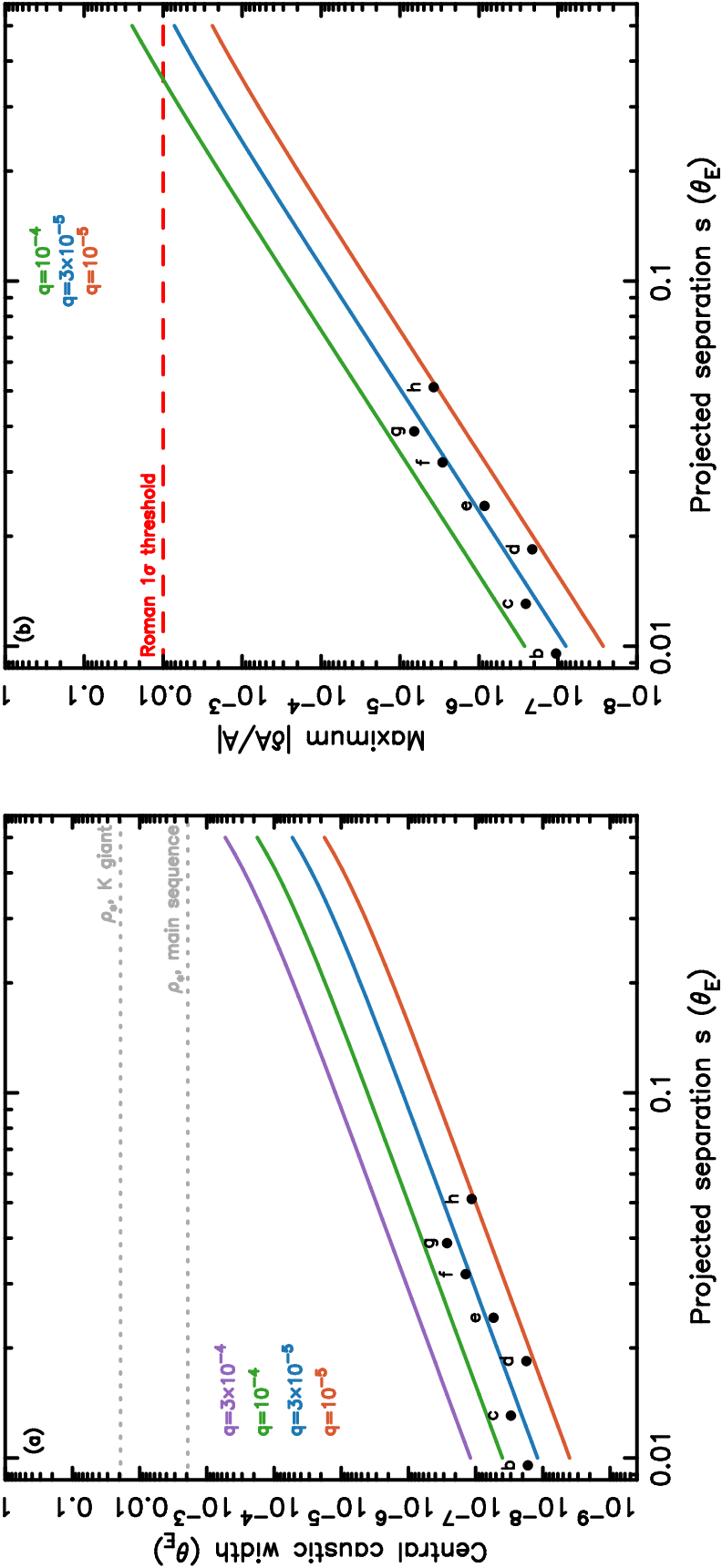}
  \caption{Central caustic properties for close planets ($s < 1$),
    both panels computed directly from the binary lens equation with
    no approximate scaling relations. (a) Width of the central caustic
    parallel to the star--planet axis, measured between the two
    on-axis cusps and determined numerically from the critical curve,
    against projected separation for several mass ratios (solid
    curves). Filled circles lettered b--h mark the seven TRAPPIST-1
    planets, and the dotted horizontal lines give the normalized
    angular radii $\rho_\star$ of a main-sequence and a K-giant bulge
    source. (b) Maximum fractional magnification deviation at $u_0 =
    0.001$ against projected separation, obtained by solving the lens
    equation along the source trajectory, with the red dashed line
    marking Roman's $1\sigma$ threshold of 10~mmag.}
  \label{fig:caustic}
\end{figure*}

Figure~\ref{fig:caustic} shows the central caustic width and the
  maximum residual as functions of projected separation. The caustic
  width scales as $q s^2$ and its shape depends only on $s$
  \citep{chung2005}. In the close regime our numerical critical curves
  give $\Delta\eta_\perp \approx \Delta\xi_\parallel$, so there is no
  favorable trajectory orientation that a survey could exploit.

%%%%%%%%%%%%%%%%%%%%%%%%%%%%%%%%%%%%%%%%%%%%%%%%%%%%%%%%%%%%%%%%%%%%

\subsection{Mass Threshold for Detectability and Dynamical Constraints}
\label{massscaling}

The combined signal of the seven planets falling seven orders of
magnitude short of detectability (Section~\ref{combined}) raises a
natural question: by what factor must the planet masses increase for
the system to become detectable via microlensing, and could the
compact resonant architecture survive at those masses?

We consider scaling every planet mass by a common factor $f$, holding
the orbital architecture fixed so that the mass ratios between planets
are preserved. Applying a common factor, rather than increasing an
individual planet, is the case most favorable to detection, since it
raises every planet's signal at once. At each $f$ we recompute
the full eight-body light curve and refit the single-lens model,
rather than extrapolating the low-mass amplitude. Because the residual
of each planet is linear in its mass (Section~\ref{scaling}), the
combined residual grows in proportion to $f$, and its $\Delta\chi^2$
as $f^2$, up to $f \sim 100$, steepening beyond as the perturbations
cease to be small. Placing all planets on-axis, the most favorable
geometry, the combined $\Delta\chi^2$ reaches the detection criterion
of 160 at $f \approx 315$ for a precision of 1~mmag, and at $f
\approx 1900$ for 10~mmag. At the first of these the most massive
planet, g, has grown to $\approx$420~$M_\oplus$, or 1.3 Jupiter
masses. At the second the planets reach up to $\sim$8 Jupiter masses
each and together amount to about 40\% of the host mass. Both
thresholds assume that a Roman epoch falls at the event peak. Other
sampling phases lower $\Delta\chi^2$ by about 1\% near $f = 315$ but
by up to 30\% near $f = 1900$, where the residual is concentrated in
a few epochs near peak, raising the second threshold by up to
$\sim$10\%. Realistic, non-aligned geometries are expected to require
still larger factors, so these values are lower bounds on the mass
amplification needed for detection.

This mass scaling, however, has immediate consequences for the
dynamical stability of the compact architecture. Following
\citet{chambers1996}, the mutual Hill radius between adjacent planets
$i$ and $i+1$ is
\begin{equation}
  R_{H,\rm mutual} = \left( \frac{M_i + M_{i+1}}{3 M_\star}
  \right)^{1/3} \frac{a_i + a_{i+1}}{2} \,,
  \label{eq:hillradius}
\end{equation}
and the mutual Hill spacing is $\Delta = (a_{i+1} - a_i) / R_{H,\rm
  mutual}$. This spacing depends only on the masses and orbits, not on
the microlensing geometry, and so is unchanged by the scale of the
lensing calculation. For the original TRAPPIST-1 system, the tightest
pair is f--g with $\Delta = 6.6$ mutual Hill radii, already below the
empirical threshold of $\Delta \approx 8$--10 for long-term stability
in generic multi-planet systems
\citep{smith2009,obertas2017,kane2020b}, and below the spacings
typical of the observed Kepler multiples \citep{fang2013}. The
TRAPPIST-1 system nonetheless survives because its mean-motion
resonance chain provides dynamical protection not captured by the Hill
criterion \citep{luger2017b,agol2021}.

Since $R_{H,\rm mutual} \propto M^{1/3}$, the Hill spacing scales as
$\Delta(f) = \Delta_{\rm base} \, f^{-1/3}$. The minimum spacing
reaches the analytic two-planet Hill stability limit of $2\sqrt{3}
\approx 3.46$ \citep{gladman1993} at $f \approx 6.8$, a factor of
about 46 below the $f \approx 315$ needed for detection at 1~mmag, and
about 280 below the $f \approx 1900$ needed at 10~mmag. At those
thresholds the minimum spacing has fallen to 0.97 and 0.53 mutual Hill
radii, respectively. Falling below this limit removes the guarantee of
Hill stability rather than proving prompt disruption, and a resonant
chain may persist somewhat further through the protection its
mean-motion resonances provide
\citep{matsumoto2012,gozdziewski2016}. Establishing the precise mass
at which a TRAPPIST-1-like chain would dissolve would require $N$-body
integrations that preserve the resonant phases, which are beyond our
scope here. Microlensing detectability and
compact dynamical stability are, for this architecture, in strong
tension.

Expanding the orbits instead of inflating the masses leaves the
dynamics unchanged. Scaling every semi-major axis by a common factor
$k$ preserves the mutual Hill spacings, since $R_{H,\rm mutual}
\propto a$ (Equation~\ref{eq:hillradius}), and the period ratios of
the resonant chain, so the stability of the architecture is
unaffected. The projected separations then grow in proportion to $k$,
and because the residual scales as $q s^3$
(Equation~\ref{eq:pertfit}), the total $\Delta\chi^2$ of the
combined signal grows as $k^6$. Computing it from the full eight-body
lens over the same window around the peak (Section~\ref{combined}), we
find that detection requires $k \approx 7$ at 1~mmag and $k \approx
15$ at 10~mmag. The outermost planet then lies at $s \approx
0.4$--0.8, with an orbital period of $\approx$380--1100~days, and the
chain extends from $\sim$0.1 to $\sim$0.5--0.9~AU. Closer to the
lensing zone the planetary caustics add further signal away from the
peak, providing the conventional route to detection. Separation
scaling therefore restores detectability only by removing the
compactness that defines the class, just as mass scaling does so only
by removing its dynamical stability.

\begin{figure*}
  \includegraphics[angle=270,width=0.9\linewidth]{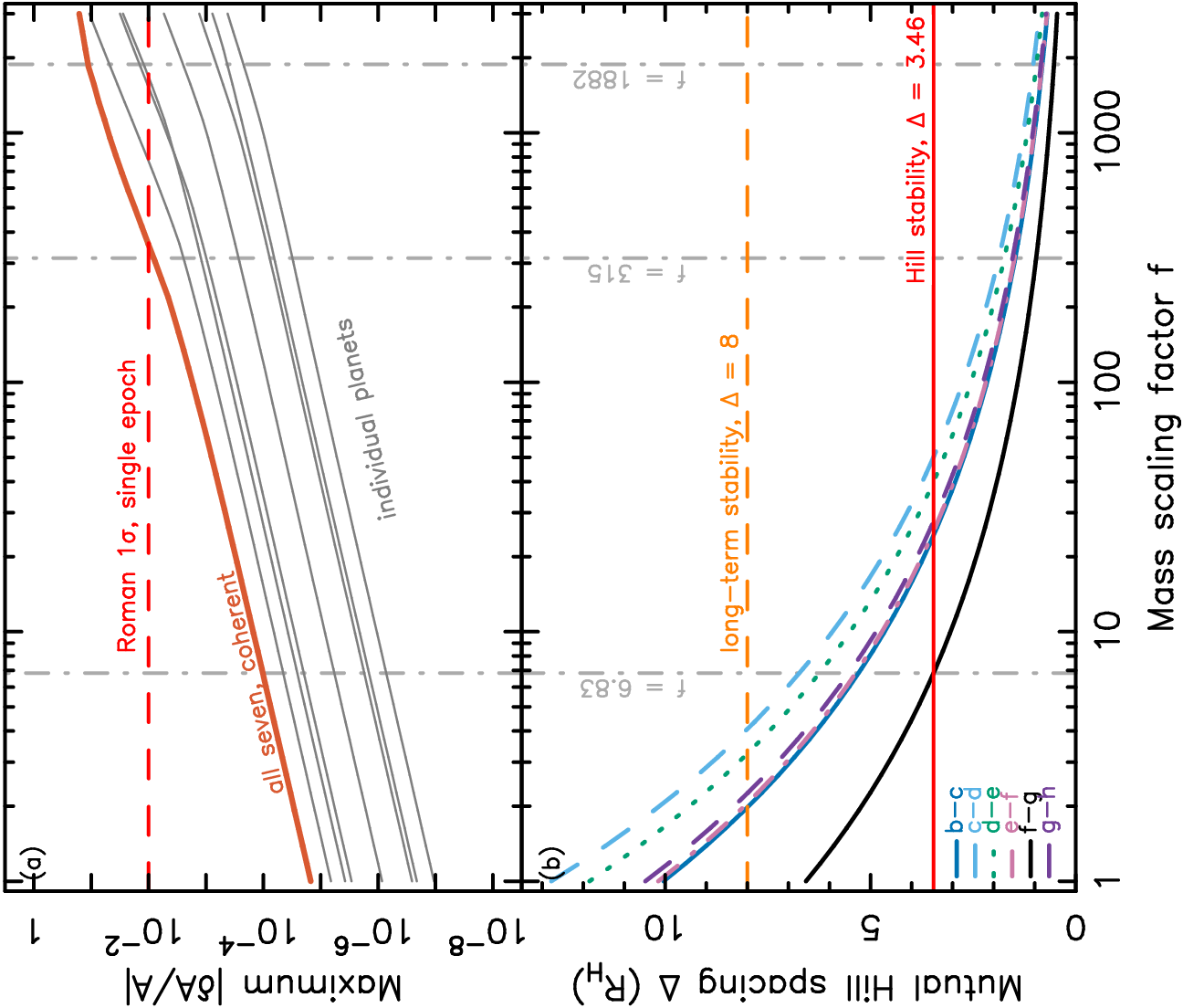}
  \caption{Detectability against dynamical stability for mass-scaled
    TRAPPIST-1 systems, in which every planet mass is multiplied by a
    common factor $f$. (a) Maximum fractional magnification deviation:
    the residual of the full eight-body lens with all seven planets
    aligned (thick orange), and the seven planets taken individually
    (thin gray). The red dashed line marks Roman's $1\sigma$
    photometric precision of 10~mmag, which is a single-epoch bound
    rather than a detection criterion; detection requires the total $\Delta\chi^2 > 160$ summed over the $\sim$250 epochs
    spanning the anomaly. The combined residual grows in proportion to
    $f$ up to $f \sim 100$ and somewhat faster beyond. (b) Mutual Hill
    spacing $\Delta$ for each of the six adjacent planet pairs,
    identified by color and line style in the legend at lower left; the f--g pair is the
    tightest at $\Delta = 6.6$ and is the first to reach the stability
    limit. The red horizontal line marks the Hill stability limit
    ($\Delta = 2\sqrt{3}$) and the orange dashed line the empirical
    long-term threshold ($\Delta = 8$). The vertical lines, common to
    both panels, mark the mass factor at which the minimum Hill
    spacing reaches the Hill stability limit ($f = 6.8$) and those at
    which the detection criterion is met for 1~mmag ($f =
    315$) and 10~mmag ($f = 1882$) precision.}
  \label{fig:massscaling}
\end{figure*}

Figure~\ref{fig:massscaling} displays this tension between
detectability and stability. The TRAPPIST-1 architecture is thus hard
to detect for a structural reason. This tension
plausibly generalizes beyond the specific TRAPPIST-1 architecture. Any
compact multi-planet system dense enough in orbital space to produce a
detectable collective microlensing signal would, by the same token,
require masses high enough to push mutual Hill spacings below the Hill
stability limit. The compactness that makes the combined signal
potentially interesting is the same compactness that enforces low
planet masses for dynamical survival.

Two simplifications regarding this analysis should be noted. First, we evaluate a single fiducial lensing geometry ($x = 0.5$),
  bounding other lens distances only approximately
  (Section~\ref{demographics}), rather than
integrating detectability over the distribution of lens and source
distances weighted by the Roman event rate, so our result is a
per-event non-detection rather than a survey-integrated occurrence
constraint. Second, we treat each event as a static configuration,
whereas the TRAPPIST-1 planets complete between one and roughly
fifteen orbits during a $22$-day Einstein crossing, so a full
treatment would follow the orbital motion through the event. The
combined signal is computed from the full eight-body lens at every
mass factor and involves no superposition of individual
perturbations. Neither simplification affects the primary result,
which falls short of the detection threshold by many orders of
magnitude, but both would need to be relaxed to convert the fiducial
non-detection reported here into a population-level occurrence
constraint. We leave that extension to future work.

%%%%%%%%%%%%%%%%%%%%%%%%%%%%%%%%%%%%%%%%%%%%%%%%%%%%%%%%%%%%%%%%%%%%

\subsection{Roman Transit Detections as an Alternative}
\label{transit}

While the microlensing channel is effectively blind to compact
systems, Roman's transit detection capability in the same bulge fields
offers a powerful alternative that directly addresses the sensitivity
gap identified in this work. The 15-minute cadence and 72-day seasons
are well matched to the short orbital periods of TRAPPIST-1-like
systems. For a UCD host, the transit depth of an Earth-sized planet is
\begin{equation}
  \delta_{\rm tr} = \left( \frac{R_\oplus}{R_\star} \right)^2 \approx
  0.59\% \,,
  \label{eq:transitdepth}
\end{equation}
adopting the TRAPPIST-1 stellar radius $R_\star = 0.119~R_\odot$
\citep{agol2021}. For TRAPPIST-1b, with a radius of $1.12~R_\oplus$,
the transit depth is 0.74\%. Whether such transits are detectable
depends on the brightness of the host, and hence on its distance,
rather than on the precision achieved for the microlensing source
stars. TRAPPIST-1 has $H = 10.72$ \citep{skrutskie2006} at a distance
of 12.47~pc \citep{vallenari2023}, so an analog at 1, 2, and 4~kpc
would have $H \approx 20.2$, 21.7, and 23.3, respectively, before
extinction. Taking the $H$ band as a proxy for the Roman filter, and
scaling Roman's precision of $\sim$10~mmag at 21~mag with a flux
dependence between the photon-limited ($\sigma \propto F^{-1/2}$) and
background-limited ($\sigma \propto F^{-1}$) cases, a single exposure
reaches 0.5--0.7\% at 1~kpc but only 3--8\% at 4~kpc. With the same
stellar radius, the 1.51-day orbit of TRAPPIST-1b yields $\sim$48
transits per 72-day season, each lasting $\sim$36 minutes, or
$\sim$2.4 Roman exposures. Six seasons therefore provide $\sim$700
in-transit exposures and a phase-folded signal-to-noise ratio of
$\approx 0.19/\sigma$, which exceeds a typical detection threshold of
$\sim$7 for $\sigma \lesssim 2.7\%$.  The detection limit for
TRAPPIST-1b thus lies at $\sim$2--4~kpc before extinction, and closer
still for the smaller outer planets, which transit less often. Transit
detections of TRAPPIST-1 analogs will therefore come predominantly
from the foreground disk, whereas the lens in our fiducial
microlensing geometry lies at 4~kpc. The two channels observe the same
fields but largely sample different distances.

\citet{tamburo2023b} performed a detailed simulation of small
transiting planet detections around mid-M and UCDs in the Roman
Galactic Bulge Time Domain Survey. They predict $1347^{+208}_{-124}$
small transiting planets, including 13 potentially habitable
terrestrial planets, from a sample of $\sim$$10^5$ sufficiently bright
cool dwarfs. This is a count of individual detected planets rather
than of resolved seven-planet analogs, but it establishes that the
transit channel is sensitive to the close-in terrestrial planets that
microlensing cannot reach, in contrast to the negligible microlensing
yield of Section~\ref{results}, underscoring the complementarity of
the two channels. The pixel-level simulations of \citet{wilson2023b}
predict $\sim$60,000--200,000 transiting planets from the survey,
including $\sim$7,000--12,000 smaller than 4~$R_\oplus$, although that
study does not address compact multi-planet systems around UCDs
specifically. While \citet{montet2017a} had previously estimated
$>$100,000 transiting planets from the Roman bulge survey, that
analysis focused primarily on hot and warm giants around FGK stars.
The \citet{tamburo2023b} result demonstrates that the sensitivity
extends to the terrestrial UCD population that microlensing cannot
access.

\begin{figure*}
  \includegraphics[angle=270,width=\linewidth]{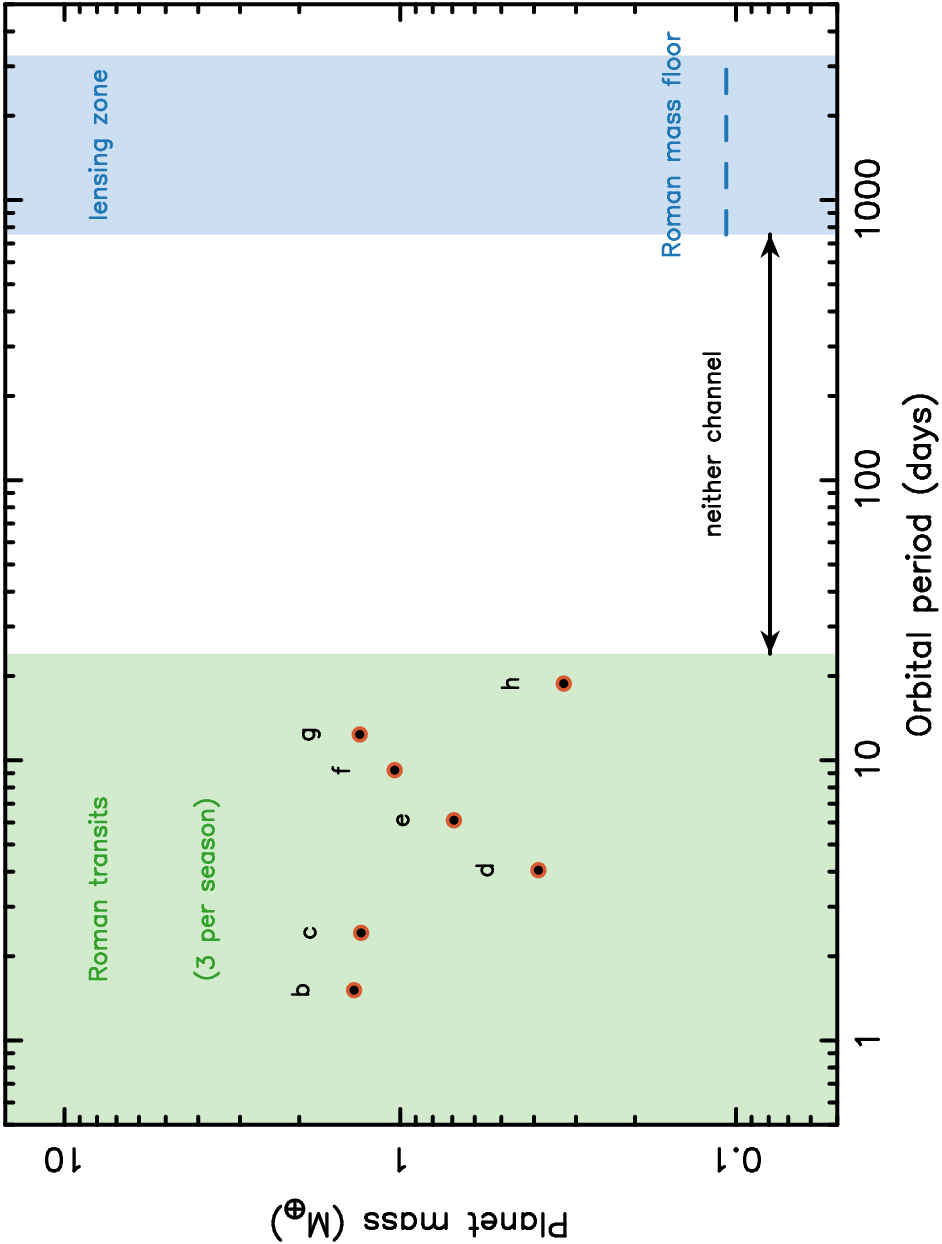}
  \caption{Sensitivity of Roman's two detection channels in the
    orbital period and planet mass plane, for a 0.09~$M_\odot$ host.
    The green band marks periods short enough to yield at least three
    transits within a 72-day observing season, $P < 24$~d. The blue
    band is the conventional lensing zone, $s = 0.6$--$1.6$ ($P =
    754$--$3282$~d). Filled circles lettered b--h mark the TRAPPIST-1
    planets, every one of which falls inside the transit band. The
    dashed line is the Mars-mass floor of Roman's microlensing
    sensitivity \citep{penny2019}, which lies below every TRAPPIST-1
    planet mass. The arrow marks the intervening range of period at
    which neither channel is optimized, representing a gap in
    sensitivity, not a hard exclusion, since detections remain
    possible there at reduced efficiency.}
  \label{fig:comparison}
\end{figure*}

Figure~\ref{fig:comparison} compares the sensitivity of Roman's two
channels in the period-mass plane. Requiring three transits within a
72-day season admits periods shorter than 24~days, which encompasses
every TRAPPIST-1 planet. The lensing zone, transformed to period
through Kepler's third law for a 0.09~$M_\odot$ host, spans 754--3282
days. The outermost planet, h, has a period of 18.8~days, a factor of
40 short of the inner edge of the lensing zone. The obstacle is
orbital period rather than planet mass, since Roman's microlensing
mass sensitivity extends to Mars-mass objects \citep{penny2019}, below
every TRAPPIST-1 planet mass. A targeted search for TRAPPIST-1 analogs
in the Roman transit sample would provide the first constraints on the
occurrence rate of compact multi-planet systems across the Galaxy,
complementing the microlensing constraints on cold planets at wider
separations.

%%%%%%%%%%%%%%%%%%%%%%%%%%%%%%%%%%%%%%%%%%%%%%%%%%%%%%%%%%%%%%%%%%%%

\section{Conclusions}
\label{conclusions}

We have presented a systematic calculation of the microlensing
detectability of compact planetary systems, using the seven-planet
TRAPPIST-1 system as a test case. By solving the binary lens equation
for each planet and evaluating the magnification perturbation against
the specifications of the Roman Space Telescope, we have shown that
for the fiducial mid-bulge geometry considered here a TRAPPIST-1
analog falls in a blind spot of the microlensing technique. The origin
of this blind spot is geometric. For a TRAPPIST-1-mass host
($0.09~M_\odot$) acting as a gravitational lens toward the Galactic
bulge, the Einstein ring radius is 1.21~AU, placing all seven planets
at projected separations of $s = 0.010$--$0.051$ in Einstein ring
units, which is deep in the close-planet regime where central caustic
perturbations are negligible. Because a close companion acts to
leading order as a single lens at the system barycenter, the
observable signal is the residual left after a single-lens model is
fitted to the light curve, which removes the unobservable
center-of-mass displacement. This residual scales as $q s^3$ and, even
for the most favorable planet (g) at the most extreme magnification
($u_0 = 0.001$), reaches only $6.6 \times 10^{-6}$, more than three
orders of magnitude below Roman's single-epoch precision of
10~mmag. The central caustics are four to five orders of magnitude
smaller than typical source angular radii, so finite-source averaging
suppresses the signal further. Summed over a full Einstein crossing at
Roman cadence, the combined $\Delta\chi^2$ of all seven planets is
$\sim$$10^{-5}$, seven orders of magnitude short of the detection
criterion of 160.

Making the system detectable would require scaling all planet masses
by a factor of $\sim$300 at 1~mmag precision, or $\sim$2000 at
10~mmag. The minimum mutual Hill spacing, however, reaches the
analytic two-planet Hill stability limit of $2\sqrt{3} \approx 3.46$
\citep{gladman1993} at a factor of only $\sim$7, removing the
guarantee of Hill stability well before the masses required for
detection are reached. At the detection thresholds, the closest
adjacent pair would be separated by less than one mutual Hill
radius. The microlensing blind spot for compact systems thus reflects
a tension between the low planet masses that keep tightly packed
architectures stable and the high masses needed for a detectable
lensing perturbation.

These results have important implications for the interpretation of
microlensing planet demographics. The non-detection of compact
terrestrial systems in current or future microlensing surveys should
not be interpreted as evidence against their prevalence. Transit
surveys indicate that compact multi-planet systems may orbit 20--44\%
of mid-M dwarfs \citep{muirhead2015,hardegreeullman2019}, and if this
extends to UCDs, such systems may be among the most common planetary
architectures in the Galaxy. For the fiducial geometry examined here,
Roman microlensing is insensitive to an individual such system, a
result consistent with the broader finding by \citet{sajadian2021c}
that Roman's HZ Earth-mass detection efficiency is $\sim$0.01\% even
for more favorable lens masses.

A comprehensive Galactic planet census will require combining
microlensing statistics for cold planets beyond the snow line with
transit statistics for hot and warm planets at short periods. Roman is
uniquely positioned to provide both, with its microlensing survey
simultaneously serving as a high-cadence transit survey of
$\sim$$10^8$ bulge stars. \citet{tamburo2023b} predict $\sim$1,350
small transiting planets around mid-M and UCDs from the same Roman
bulge fields, demonstrating that the transit channel can access
precisely the compact terrestrial population that the microlensing
channel cannot. A dedicated search for TRAPPIST-1 analogs in the Roman
transit data would constitute the first probe of compact terrestrial
system occurrence across the Galaxy, filling the demographic gap
quantified by this work.

%%%%%%%%%%%%%%%%%%%%%%%%%%%%%%%%%%%%%%%%%%%%%%%%%%%%%%%%%%%%%%%%%%%%

\section*{Acknowledgements}

The author would like to thank the anonymous referee, whose feedback
helped to improve the manuscript. The results reported herein
benefited from collaborations and/or information exchange within
NASA's Nexus for Exoplanet System Science (NExSS) research
coordination network sponsored by NASA's Science Mission Directorate.

%%%%%%%%%%%%%%%%%%%%%%%%%%%%%%%%%%%%%%%%%%%%%%%%%%%%%%%%%%%%%%%%%%%%

%\bibliographystyle{aasjournal}
%\bibliography{/data/skane/latex/styles/references}

%%%%%%%%%%%%%%%%%%%%%%%%%%%%%%%%%%%%%%%%%%%%%%%%%%%%%%%%%%%%%%%%%%%%

\end{document}